L.V.Grunskaya[1], V.V.Isakevich[2], D.V.Isakevich[3], L.T.Sushkova[4]

# DETECTING COMPONENTS SPECTRALLY LOCALIZED AT ASTROPHYSICAL PROCESS FREQUENCIES IN TIME SERIES OF THE ELECTRIC FIELD VERTICAL COMPONENT OF THE EARTH ATMOSPHERE BOUNDARY LAYER[5][6].

**Summary**

Signal eigenvectors and components analyser (RF Utility model patent 116242) was used to explore the time-series of the electric field vertical component $E_z$ in the Earth atmosphere boundary layer. There have been detected non-coherent complex-periodic components localized at the frequencies of gravity-wave impact of binary stars and at the frequency of axion-photon interaction. These components cannot be detected by means of quadrature scheme of spectral analysis and have RMS values from 0.05 V/m to 0.5 V/m at binary stars gravity-wave impact frequencies and from 0.7 V/m to 2.7 V/m at axion-photon interaction frequency. The probability of the eigenvectors localization at the binary stars gravity-wave frequencies being random is in the range $1.2 \cdot 10^{-4} - 9.8 \cdot 10^{-3}$ for each of the analysed time series. This probability estimated for all the time series is not more than $10^{-9}$. The probability of the eigenvectors localization at the axion-photon frequency being random is not more than 0.06. For all the analysed $E_z$ time series there have been detected components that were spectrally localized at the total and difference frequencies of binary stars gravity-wave impact and of axion-photon interaction and had RMS values from 0.05 V/m to 0.59 V/m. This fact implies that the components localized at binary stars gravity-wave impact frequencies have amplitude modulation of axion-photon interaction frequency. The false detection probability of this effect for each of the analysed time series is $1.9 \cdot 10^{-6} - 4.3 \cdot 10^{-3}$. For all the time series the false detection probability is not more than $1.2 \cdot 10^{-15}$. It was also demonstrated that the axion-photon interaction frequency modulates the amplitude of the components spectrally localized at lunar and solar tides frequencies.

**Introduction**

Exploring electromagnetic field of the Earth atmosphere boundary layer [1] in the range of frequencies close to lunar and solar tides resulted in detecting the non-coherent components in the electric and geomagnetic field with their spectrum localized at the frequencies of relativistic binary stars gravity-wave emission. The reliability of this detection is confirmed by the analysis of the Earth electric field long-term time series got at several stations in various regions of Russia.

---

[1] E-mail: grunsk@vlsu.ru
[2] E-mail: businesssoftservice@gmail.com
[3] E-mail: voiceofhope@yandex.ru
[4] E-mail: ludm@vlsu.ru

[5] The work was carried out with the support of the State Assignment 2014/13,2871 and the RFRF Grant № 14-07-97510/14
[6] Calculations were made using the GNU Octave software, CeCILL Scilab, GNU G95

During a long time range (1997-2012) of accumulating experimental data we've created a spatially separated detecting complex for monitoring the Earth atmosphere boundary layer electromagnetic fields in the infra-low frequency range (ILF)[2]. At the same time we've improved the time series processing method. Thus appeared an innovation decision named "Signal eigenvectors and components analyser" (SEV&CA) [3].

Indirect detection of periodic components localized at the frequencies of gravity-wave (GW) impacts from astrophysical sources by using the induced electromagnetic fields in the Earth-atmosphere resonator is an actual problem, both in theoretical and experimental aspect. The theoretic fundamentals of quasi-static electromagnetic fields and of gravity emissions were made by the works of D.Bocaletti, V.L.Ginzburg, Y.B.Zel'dovich [4-6]. It was demonstrated that when a periodic gravity emission affects a magnetic field it makes the latter change with the frequency of the gravity emission. These pioneer works caused a surge of interest both in our country and abroad. Theory stated that the mentioned effect should result in the additive component of electromagnetic potential. The component strongly depends on the properties of the environment containing the original magnetic field. This effect is extremely weak in vacuum, however sometimes an abnormal increase in the response is observed [7-9]. The idea of considering the Earth as a GW emissions detector has arisen a relatively long time ago. The review made by V.K.Malyukov and V.N.Rudenko [10] refers to the works of J.Veber and T.Dyson written in the 1960s where the authors studied the possibility of seismographic detecting of GW emissions. The problem of direct seismographic detection of the Earth mechanical oscillations caused by GW emissions still looks utopian because of geological interference. However, studying the Earth electromagnetic fields in the infra-low frequency range (ILF), as it will be demonstrated further, makes it possible to detect the GW emission with high reliability.

As it was forecasted theoretically, monitoring the ILF variations of the Earth electric and geomagnetic field can result in detecting the traces of axion-photon interactions [11]. The authors of the axion-photon interaction research program [11] basing on the monitoring data of the Earth electromagnetic field ILF variations proposed a hypothesis according to which in dark matter aggregations close to the Earth relict axions have heterogenic distribution. Theoretic research tells that axion-photon interactions in the electric field of the Earth atmosphere boundary layer have to be detected at the frequency close to $5 \cdot 10^{-6}$ Hz [11].

This work concerns the problem of detecting those components in the vertical electric field of the Earth atmosphere boundary layer that are spectrally localized at the frequencies of gravity-wave emissions of astrophysical sources (binary star systems – BSS) and at the frequency of axion-photon interaction.

Like it was done in the previous work the time series (TS) were processed by the signal eigenvectors and components analyser (SEV&CA [3]) using the following parameters: analysis range M = 1000 counts, discretization time Δt=1 hr. The analysis was carried out for the following four long-term TS of the electric field vertical component $E_z$ in the Earth atmosphere boundary layer: the data from the testing ground of the VlSU Department of General and Applied Physics (2003--2009), The data from geophysical observatories of Dusheti (1976–1980), Voyeikovo (1966–1995), Verkhneye Dubrovo (1974–1995). For detecting spectrally localized components the eigenvectors (EV) spectral analysis was used, selection was performed according to the coherence index (CI) equal to the fast Fourier transform (FFT) amplitude at the sought frequency taken relative to FFT amplitude mean value. CI is similar to "signal-noise" ratio.

**Main part**

The sources taken for the analysis are described in table 1. We selected those sources whose frequency differs significantly from that of lunar tides and is not more than a half of discretization frequency. In the first column there is a name of the source, in the second column – the corresponding frequency. Corresponding to this source the fifth column of the table contains rounded to integer ratio of the difference between the analysed source frequency and the nearest frequency of other sources to the frequency resolution[6]. The name and value of the nearest frequency are in the third and fourth columns correspondingly. The values in the fifth column of table 1 show the resolution reserve of SEV&CA for eigenvectors spectral analysis when M=1000. So, when detecting components at frequencies from table 1 the value of spectral resolution is enough for reliable frequency separation.

Like it was in the previous work [1] the components of the considered frequencies in the analysed time series appeared undetectable by means of a standard quadrature scheme of time series spectral analysis, this being a result of the components incoherence in the analysis time-range. As an example figure 1 demonstrates the dependence of spectral component amplitude estimate vs. analysis time-range value using a quadrature circuit or 3 frequencies from table 1. The plots confirm the absence of coherent components at the considered frequencies. Such dependencies are similar for all frequencies from table 1 and for all the four time series.

**Table 1 Sources and their frequencies**

| Source considered | | Source nearest by frequency | | $\dfrac{|f_{cons} - f_{nearest}|}{\Delta f}$ |
|---|---|---|---|---|
| Name | $f_{cons}$, $10^{-5}$ Hz | Name | $f_{nearest}$, $10^{-5}$ Hz | |
| Axion-photon interaction | 0.5 | Tide $O_1$ | 1.075921027515 | 20 |
| J1012+5307 | 3.828211138105 | Tide $S_3$ | 3.472222 | 12 |
| J1537+1155 | 5.501805538757 | J1959+2048 | 6.06253904577 | 20 |
| J1959+2048 | 6.06253904577 | J1537+1155 | 5.501805538757 | 20 |
| J2130+1210 | 6.904082103431 | J1915+1606 | 7.16666560145 | 9 |
| J1915+1606 | 7.16666560145 | J2130+1210 | 6.904082103431 | 9 |

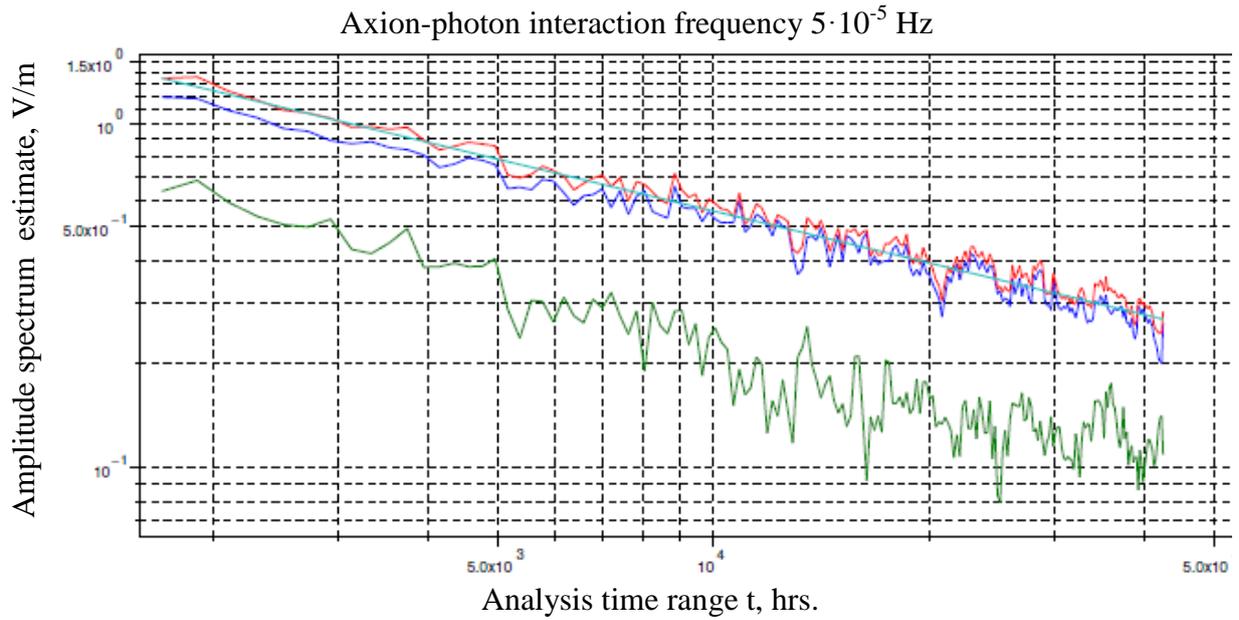

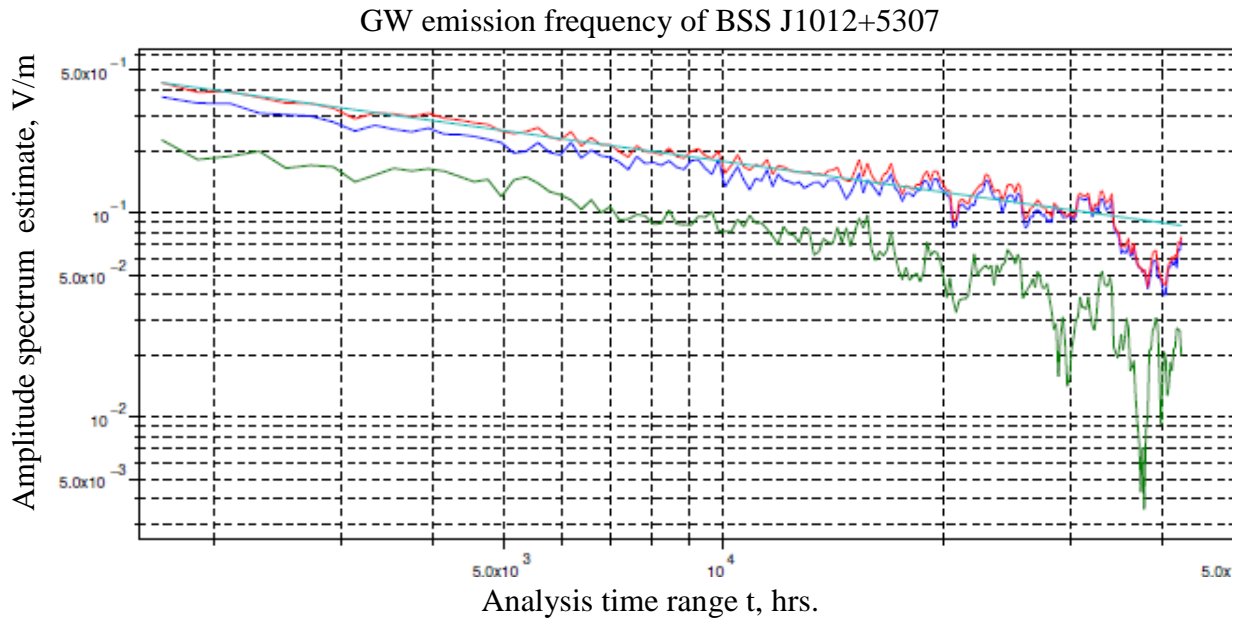

Fig.1. Amplitude estimate of the spectral component for time series at Voyeikovo station vs. time range value when using quadrature scheme. Upper curve corresponds to RMS value, middle curve – sample mean, lower curve — sample variance. The straight line $\sim 1/\sqrt{t}$ where $t$ is the time-range length.

Figures 2 – 6 demonstrate some localized at considered frequencies eigenvectors and their amplitude spectra got by FFT.

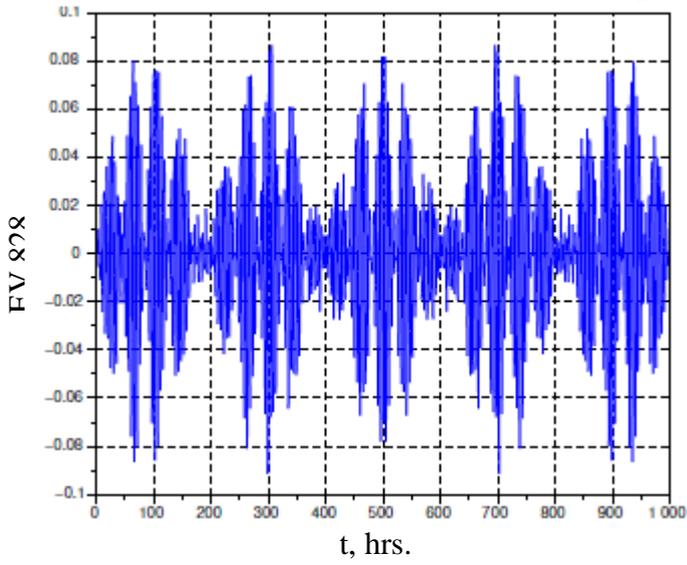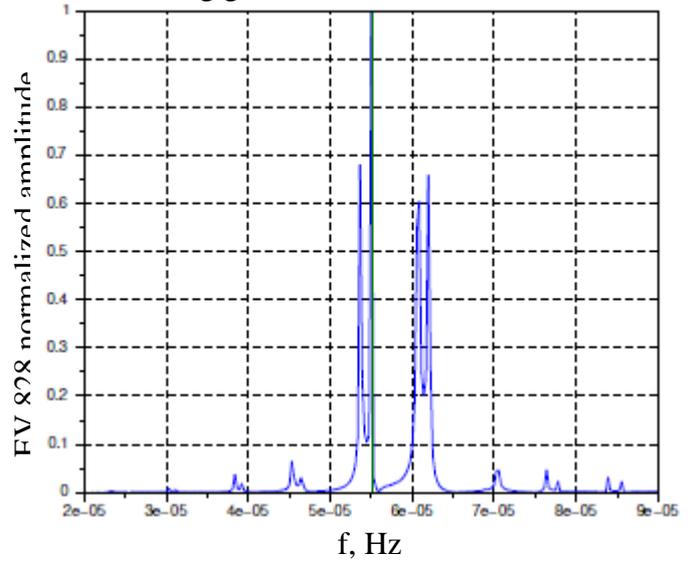

J1537+1155, E$_z$, data from VSU testing ground

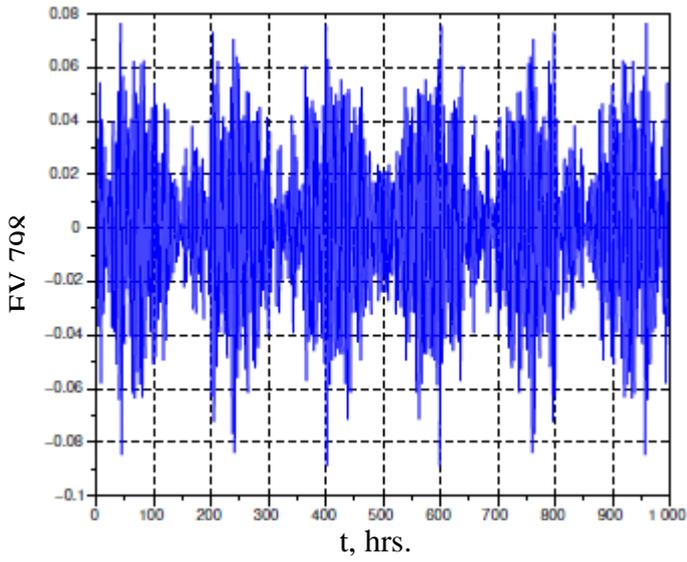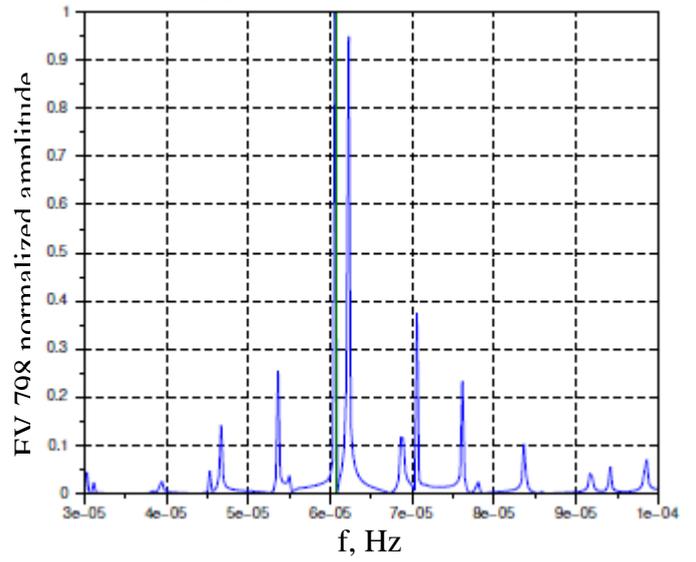

J1959+2048, E$_z$, data from VSU testing ground

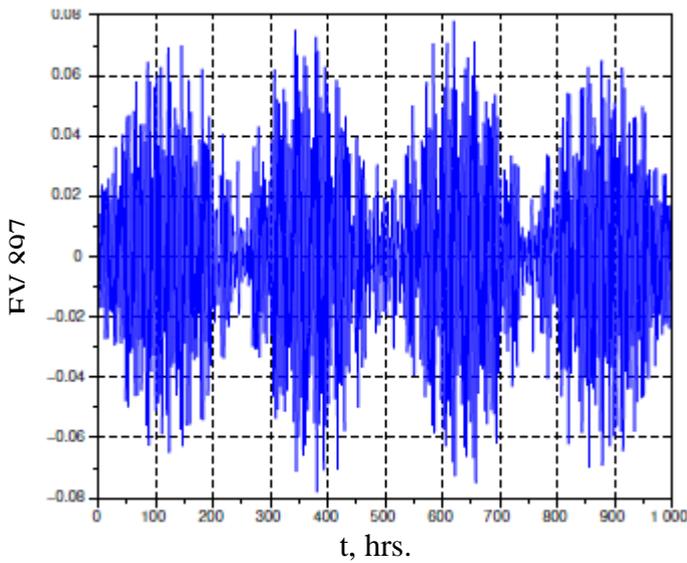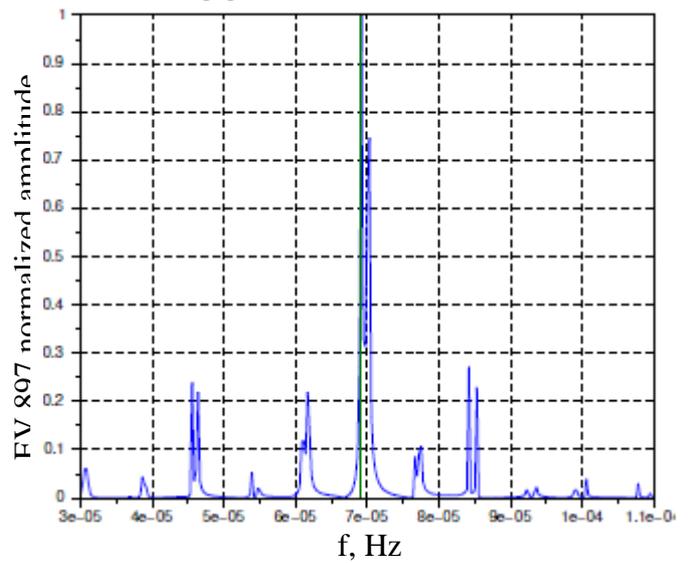

J2130+1210, E$_z$, data from VSU testing ground

Fig.2. Eigenvectors (left) localized at BSS GW emission and their normalized amplitude spectra (right). Continuous vertical line at the spectrum plots relates to the considered frequency.

J1012+5307, E$_z$, data from Voyeikovo station

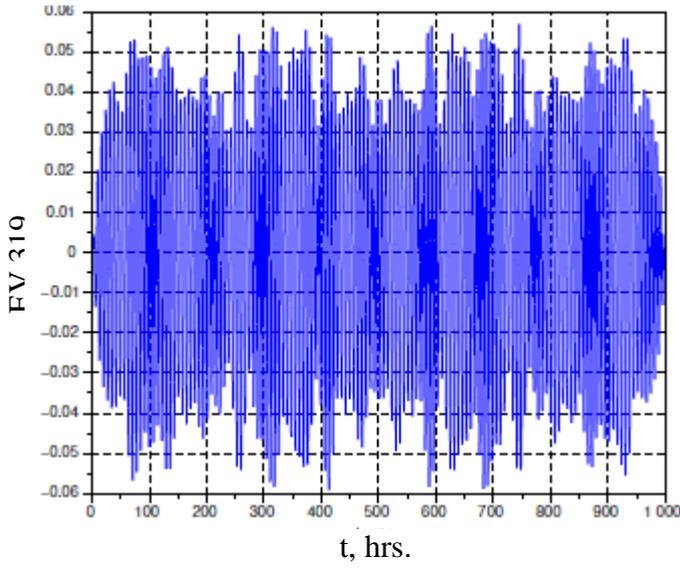 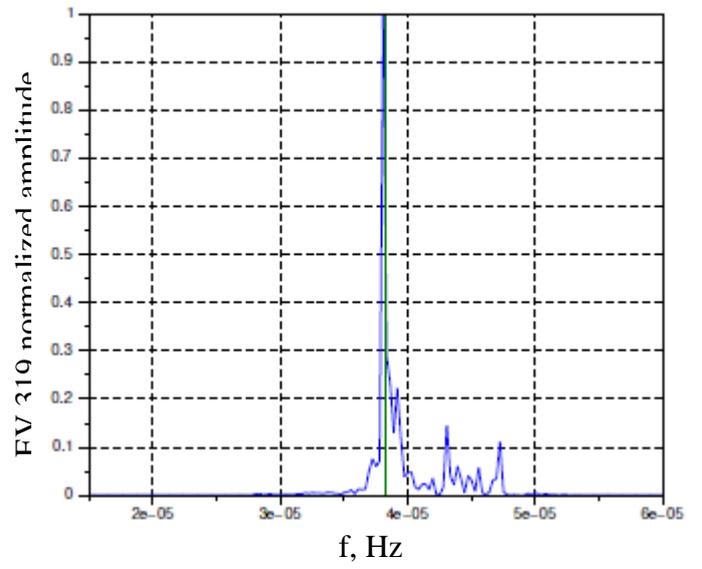

J1915+1606, E$_z$, data from Voyeikovo station

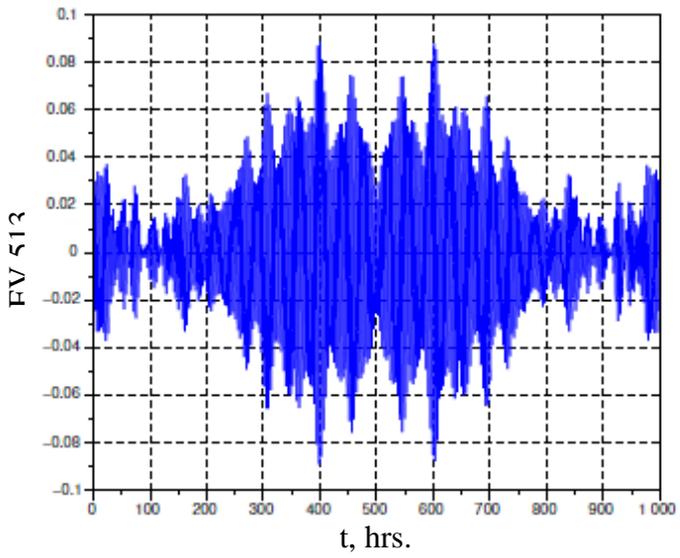 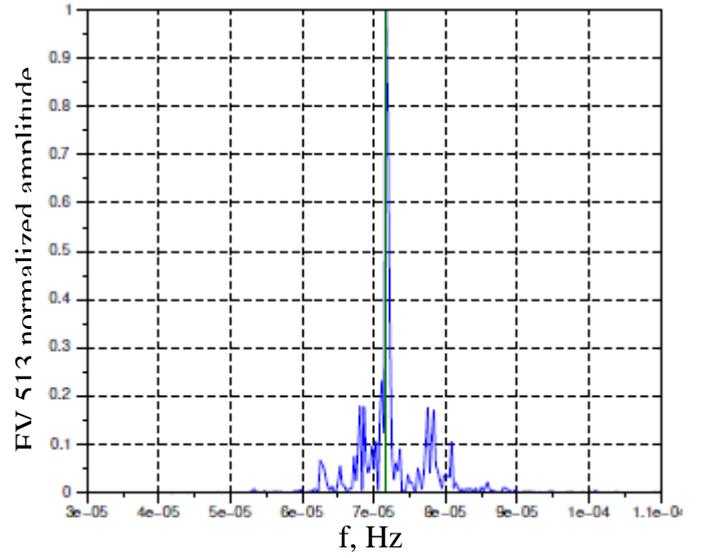

J1537+1155, E$_z$, data from Voyeikovo station

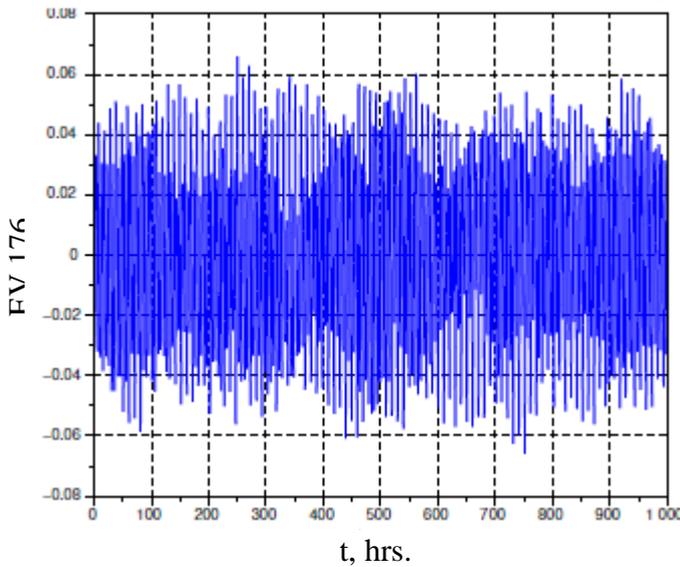 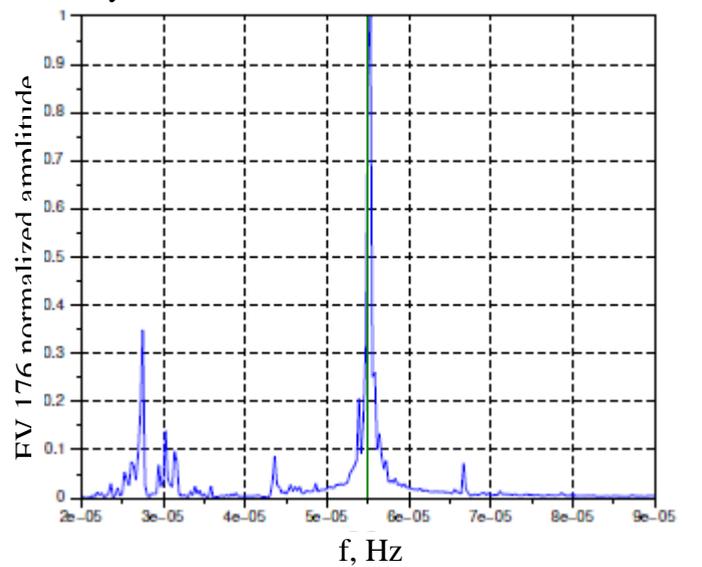

Fig.3. Eigenvectors (left) localized at BSS GW emission and their normalized amplitude spectra (right). Continuous vertical line at the spectrum plots relates to the considered frequency.

J1012+5307, E$_z$, data from Dusheti station

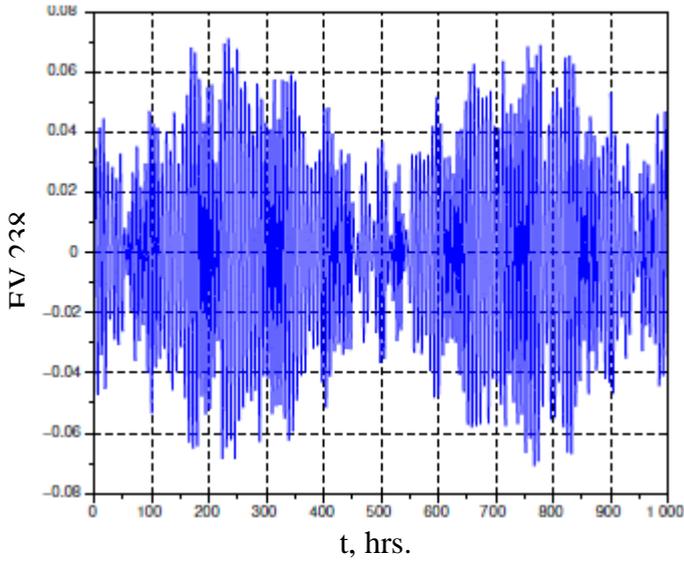 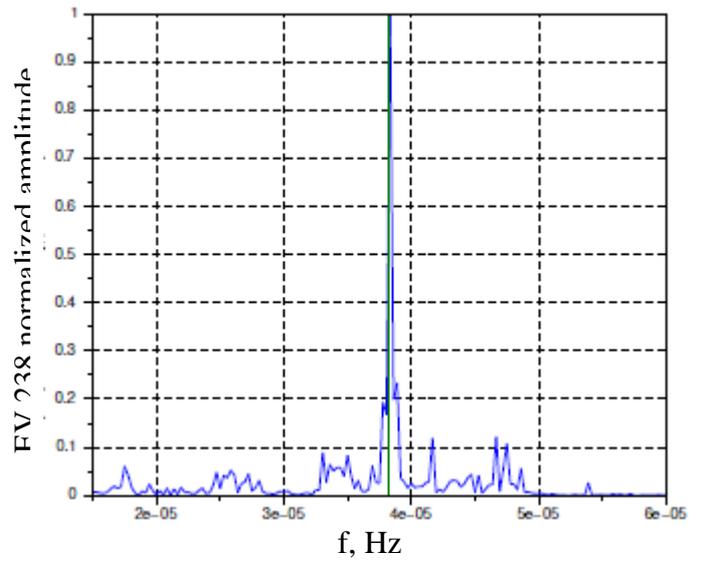

J1959+2048, E$_z$, data from Dusheti station

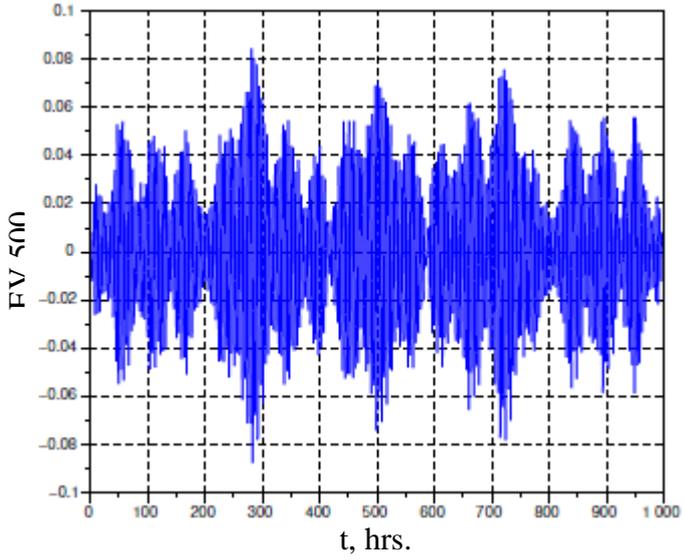 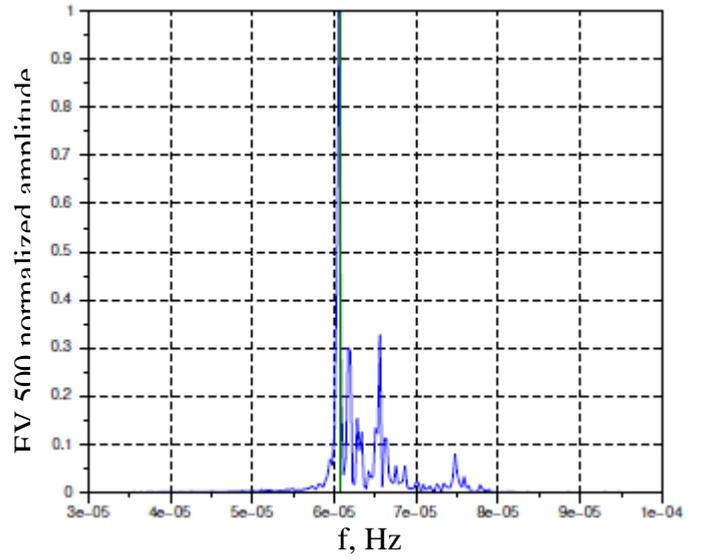

J2130+1210, E$_z$, data from Dusheti station

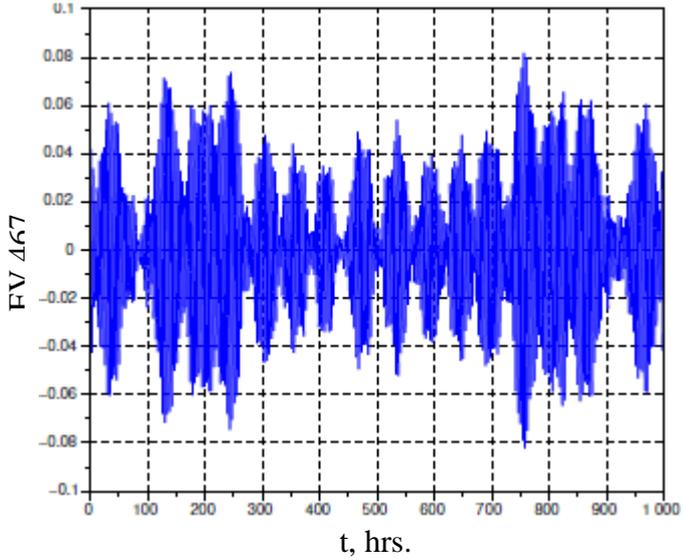 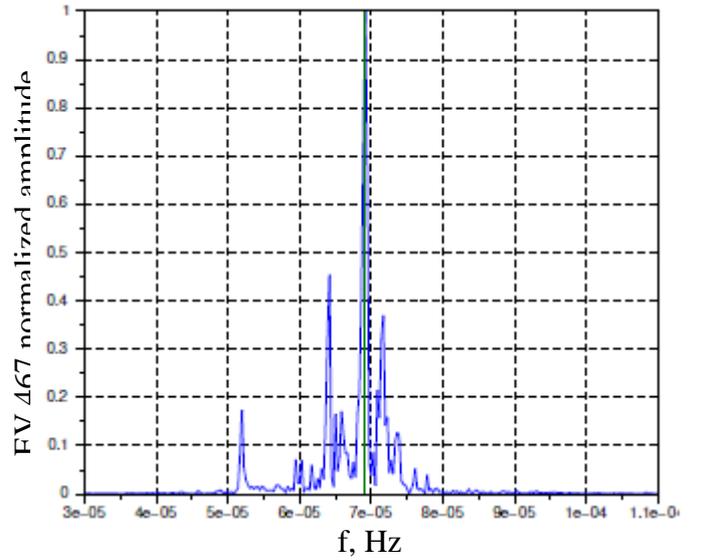

Fig.4. Eigenvectors (left) localized at BSS GW emission and their normalized amplitude spectra (right). Continuous vertical line at the spectrum plots relates to the considered frequency.

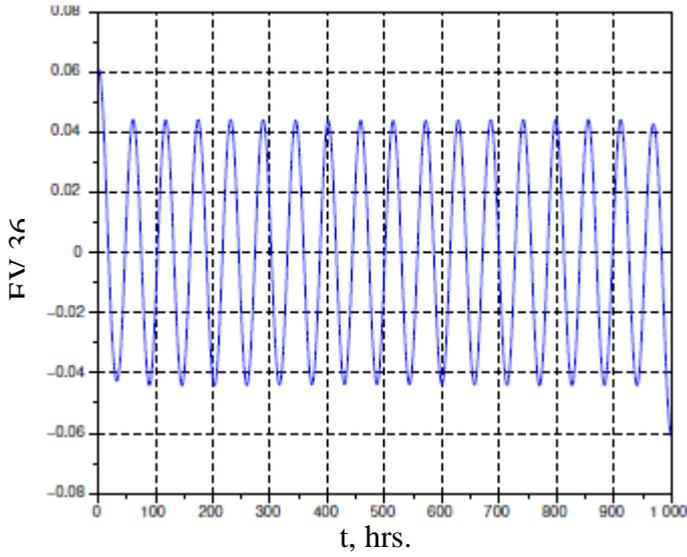 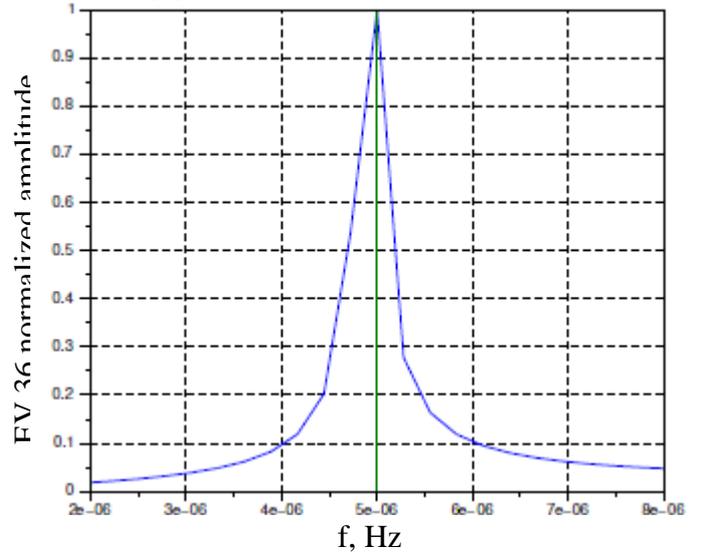

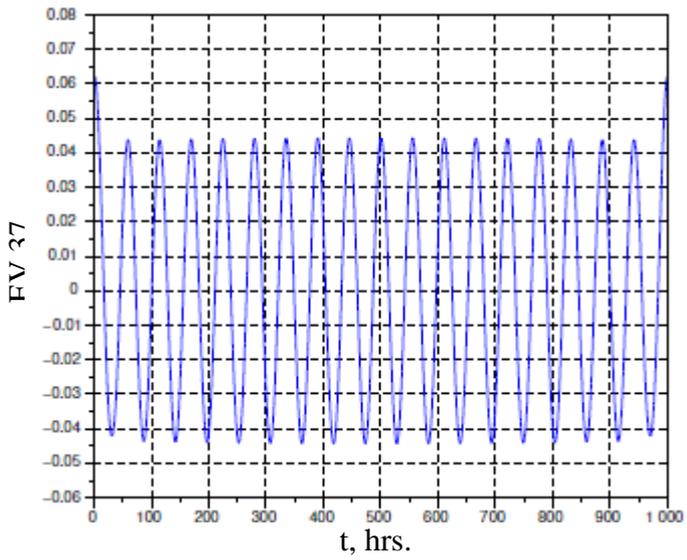 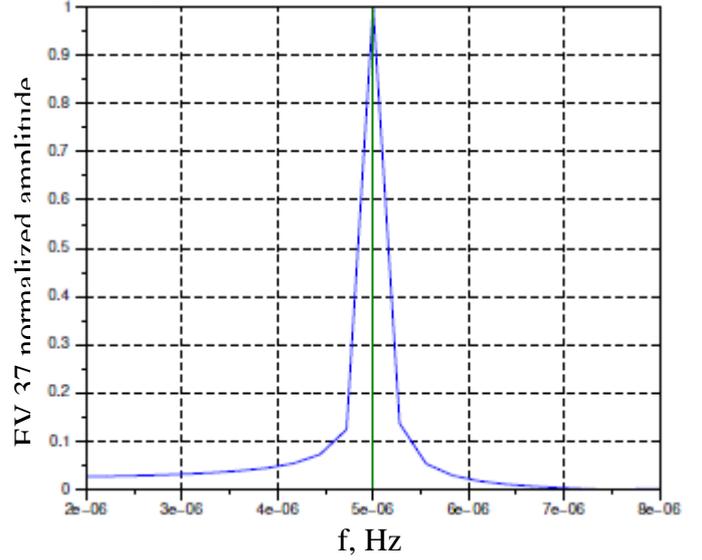

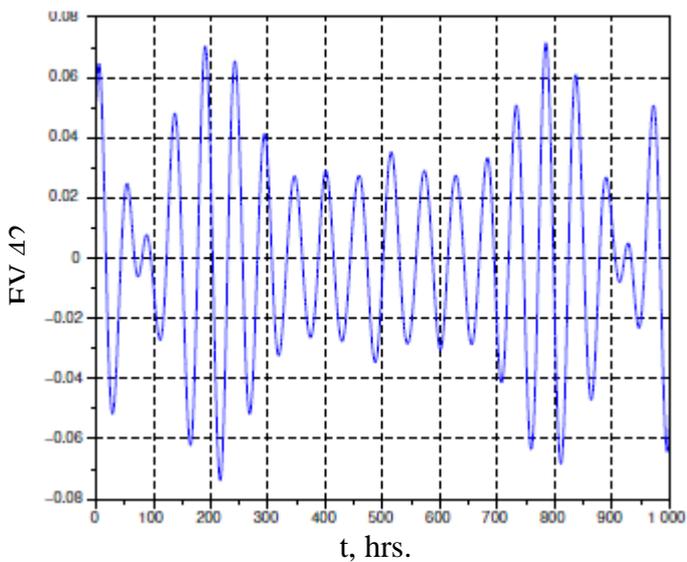 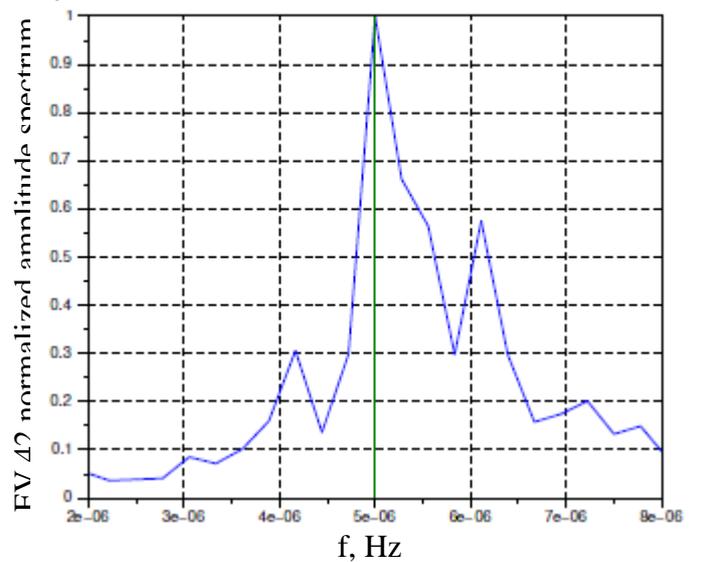

Fig.5. Eigenvectors (left) localized at axion-photon interaction frequency and their normalized amplitude spectra (right). Continuous vertical line at the spectrum plots relates to the considered frequency.

API, E$_z$, data from Voyeikovo station

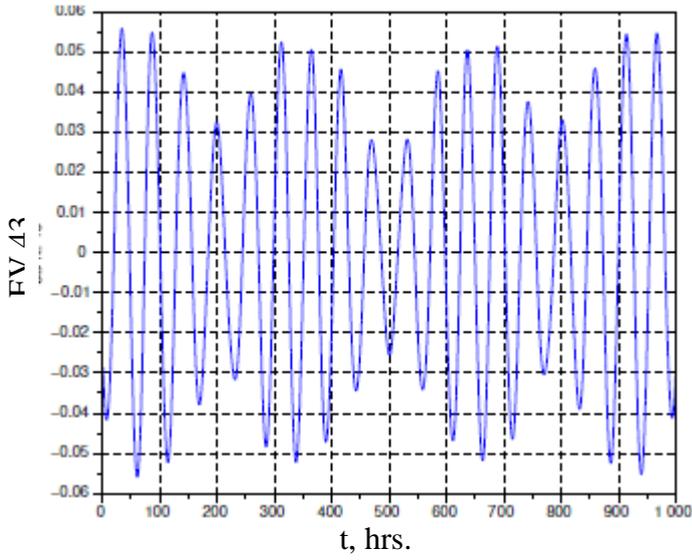 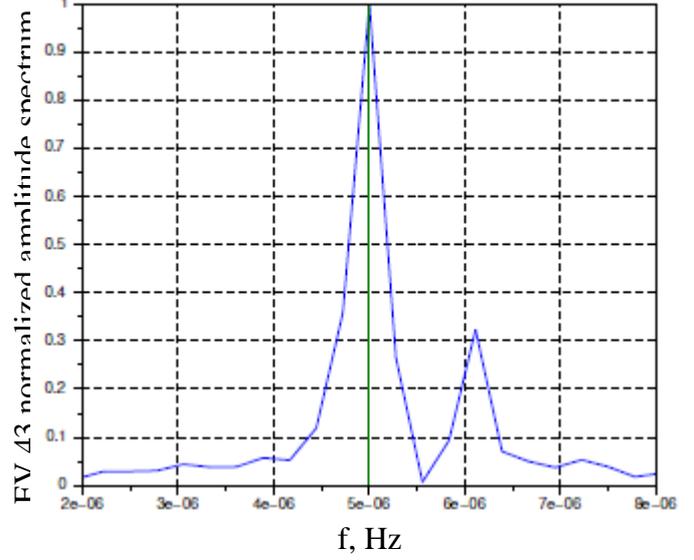

API, E$_z$, data from Verkhneye Dubrovo station

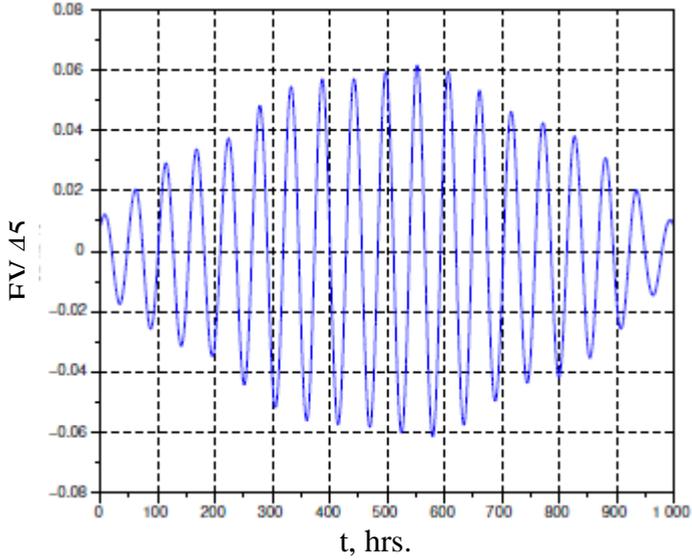 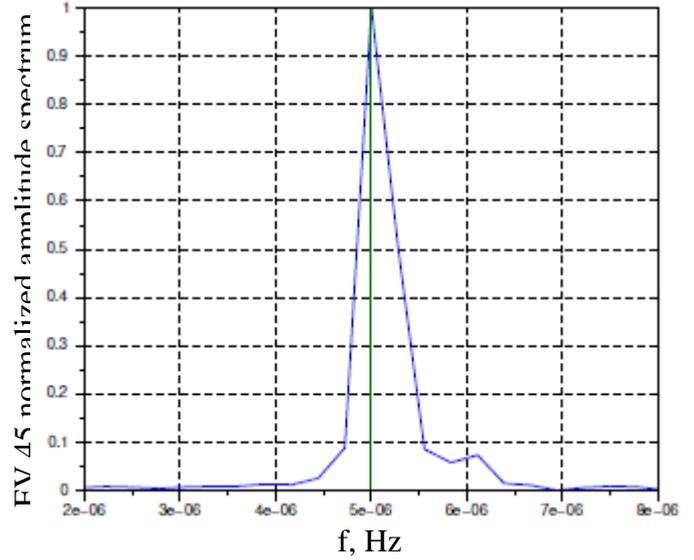

API, E$_z$, data from Verkhneye Dubrovo station

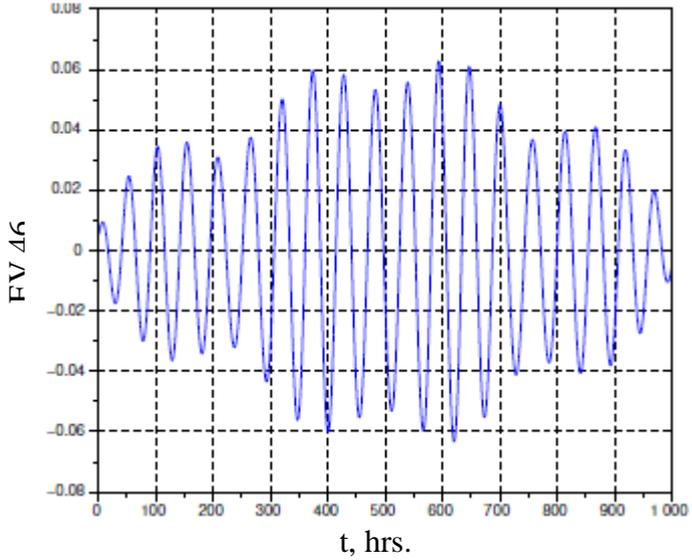 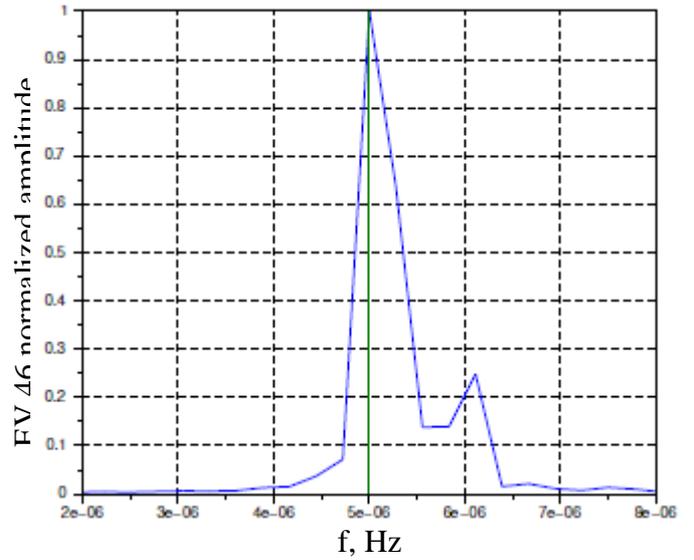

Fig.6. Eigenvectors (left) localized at axion-photon interaction frequency and their normalized amplitude spectra (right). Continuous vertical line at the spectrum plots relates to the considered frequency.

In figures 2 to 6 there are shown some eigen vectors (EV) and their amplitude spectra when using FFT localized at the revealed frequencies.

Tables 2 and 3 contain summary information concerning the behavior of eigenvectors (EV) having spectrum maximums at the frequencies given in table 1 as well as data about CI and RMS values got by using the method mentioned in the previous work and described in details in [12]. The penultimate column of table 2 contains numbers of the selected EV coherence indexes exceeding the CI median value (column 4). The data for the considered frequency have been got as a result of calculating experiment 1 described in the previous work [1]. The number of CI exceeding median value allows to make an upper estimate of "false alarm" probability $P_{fa}$ when detecting periodic components having frequencies from table 1 both for each observation station (this probability is shown in the corresponding section of table 2) and for all the observed time series at all the stations (in the lower section of the table).

**Table 2. Coherence indexes (CI) and RMS values of eigen vectors spectrally localized at the frequencies of BSS GW emission (the results of observed $E_z$ time series analysis).**

| Source | EV eigen vectors | EV CI | CI median | Median exceeding | Amplitude RMS value, V/m |
|---|---|---|---|---|---|
| VSU testing ground, $P_{fa}(9, 10)=9.8 \cdot 10^{-3}$ | | | | | |
| J1012+5307 | 780; 779 | 45.2; 53.1 | 34.7 | 2 of 2 | 0.092; 0.092 |
| J1537+1155 | 829; 828 | 64.2; 61.7 | 30.4 | 2 of 2 | 0.077; 0.077 |
| J1959+2048 | 799; 798 | 58.1; 61.9 | 24.0 | 2 of 2 | 0.088; 0.088 |
| J2130+1210 | 898; 897 | 68.2; 61.9 | 27.3 | 2 of 2 | 0.055; 0.055 |
| J1915+1606 | 523; 521 | 36.3; 26.0 | 28.8 | 1 of 2 | 0.16; 0.16 |
| Voyeikovo, $P_{fa}(13, 13)=1.2 \cdot 10^{-4}$ | | | | | |
| J1012+5307 | 273 | 70.5 | 34.7 | 1 of 1 | 0.44 |
| J1537+1155 | 382; 435; 436 | 64.6; 58.0; 71.6 | 30.4 | 3 of 3 | 0.36; 0.33; 0.33 |
| J1959+2048 | 399; 400; 427 | 72.1; 76.9; 34.0 | 24.0 | 3 of 3 | 0.34; 0.34; 0.33 |
| J2130+1210 | 568; 491 | 38.6; 28.3 | 27.3 | 2 of 2 | 0.29; 0.30 |
| J1915+1606 | 514; 513; 512; 478 | 36.8; 84.1; 44.5; 38.0 | 28.8 | 4 of 4 | 0.30; 0.30; 0.30; 0.31 |
| Dusheti, $P_{fa}(11, 12)=2.9 \cdot 10^{-3}$ | | | | | |
| J1012+5307 | 239; 238 | 98.5; 98.3 | 34.7 | 2 of 2 | 0.51; 0.51 |
| J1537+1155 | 376; 375 | 37.8; 41.4 | 30.4 | 2 of 2 | 0.44; 0.44 |
| J1959+2048 | 845; 825; 824 | & 31.1; 57.3; 40.5 | 24.0 | 3 of 3 | 0.35; 0.36; 0.36 |
| J2130+1210 | 340; 339 | 58.3; 55.6 | 27.3 | 2 of 2 | 0.45; 0.46 |
| J1915+1606 | 440; 437; 434 | & 20.0; 29.0; 29.4 | 28.8 | 2 of 3 | 0.43; 0.43; 0.43 |
| Verkhneye Dubrovo, $P_{fa}(8, 9)=3.1 \cdot 10^{-2}$ | | | | | |
| J1012+5307 | 257 | 100.0 | 34.7 | 1 of 1 | 0.41 |
| J1537+1155 | 404; 403 | 171.0; 165.0 | 30.4 | 2 of 2 | 0.31; 0.31 |
| J1959+2048 | 500; 499 | 95.2; 74.2 | 24.0 | 2 of 2 | 0.27; 0.27 |
| J2130+1210 | 467; 466 | 55.1; 47.8 | 27.3 | 2 of 2 | 0.28; 0.28 |
| J1915+1606 | 547; 544 | 26.2; 31.2 | 28.8 | 1 of 2 | 0.26; 0.26 |
| For all the 4 time series $P_{fa}(41,44)=7.5 \cdot 10^{-10}$ | | | | | |

The false alarm probability $P_{fa}$ was estimated according to the method described in [1]. $P_{fa}$ varies from $P_{fa}(13, 13)=1.2 \cdot 10^{-4}$ (Voyeikovo) to $P_{fa}(8, 9)=3.1 \cdot 10^{-2}$ (Verkhneye Dubrovo). For all the observed time series the probability of false detecting components having BSS GW emission frequencies $P_{fa}(41, 44) = 7.5 \cdot 10^{-10}$.

Table 3 contains similar data for the axion-photon frequency. Here the estimate of false detection probability for spectrally localized components is not more than $5.5 \cdot 10^{-2}$ for all the observed $E_z$ time series. This estimate can be improved only by including in consideration new time series of $E_z$ observations.

**Table 3. Coherence indexes and RMS values of eigen vectors spectrally localized at axion-photon interaction frequency (the results of observed $E_z$ time series analysis)**

| Station | EV | EV CI | Amplitude RMS value, V/m | CI median | Median exceeding | $P_{fa}$ |
|---|---|---|---|---|---|---|
| VSU testing ground | 36; 37 | 73.6; 248.5 | 2.7; 2.6 | 63.8 | 6 of 7 | $5.5 \cdot 10^{-2}$ |
| Dusheti | 63 | 133.3 | 0.71 | | | |
| Verkhneye Dubrovo | 45; 46 | 238; 185.2 | 1.3; 1.3 | | | |
| Voyeikovo | 42; 43 | & 50.2; 151.8 | 1.3; 1.3 | | | |

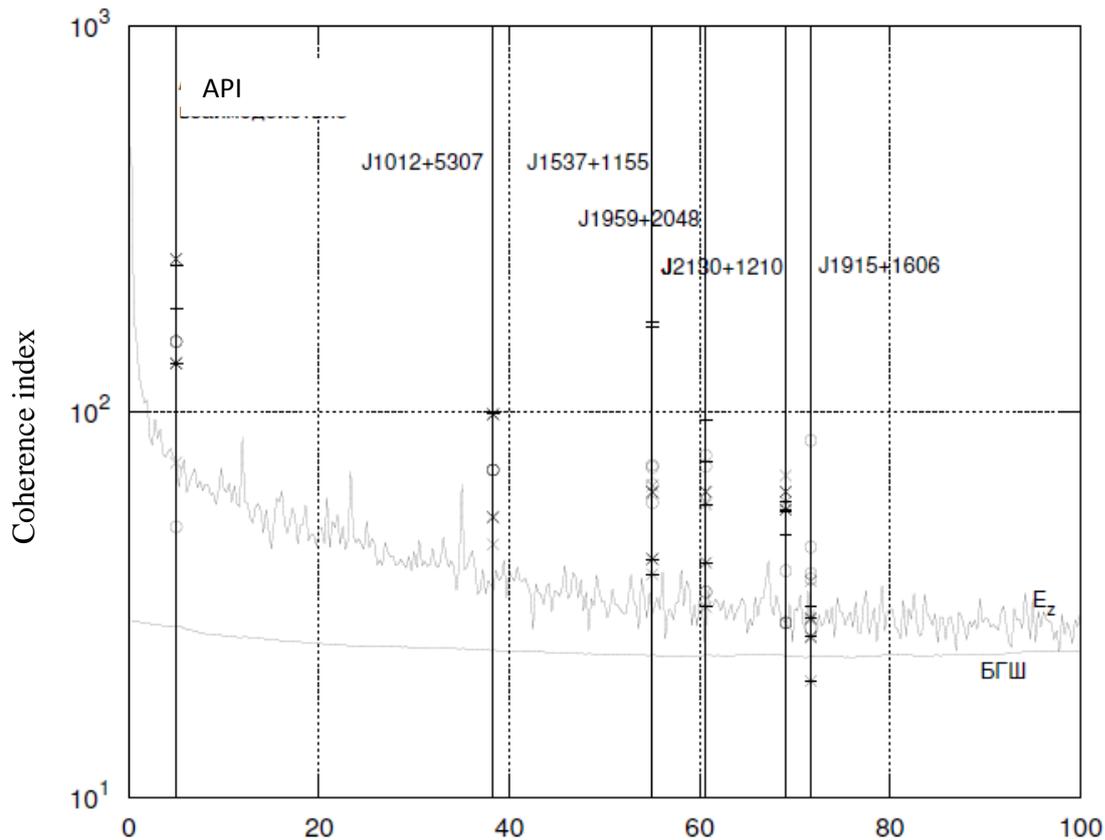

Fig.7. CI vs. position of EV amplitude spectrum maximum. The result of calculating experiment for $E_z$ TS segments (upper curve) and for Gaussian white noise (БГШ) segments of the same length (lower curve) in comparison to CI of $E_z$ TS eigenvectors spectrally localized at BSS GW emission frequencies and API frequency. Vertical lines show considered frequencies × — VSU; * — Dusheti; ° — Voyeikovo + Verkhneye Dubrovo.

The results have been got using the method given in [1]. Coherence indexes are calculated for the frequency range corresponding to API and BSS GW emission.

Figure 7 illustrates selected eigenvectors CI values compared to median CI values in the frequency range of API and BSS GW emission. The results have been got during the calculating experiment done according to the method given in [1].

It was mentioned in SEV&CA patent specification [3] that if the considered TS contains some spectrally localized harmonic component whose amplitude is modulated by other harmonic TS (other spectrally localized component) then among the TS eigenvectors should be those having the amplitude spectrum localized at total and difference frequencies of the two TS – the modulated TS and the modulating one (total and difference frequencies of their spectral localization).
This conclusion motivated us to search for those EV whose amplitude spectra are localized at total and difference frequencies of GW impact and of API. Table 5, similar to table 1, presents total and difference frequencies to be considered in order to confirm the effect of API-frequency amplitude modulation of $E_z$ time series components having frequencies of BSS GW impact. The frequencies in the first column are named after BSS adding "+" for total frequencies or "-" for difference frequencies. Similar to table 1, considered frequencies are compared to the nearest ones in order to assess the possibility of their EV separation using FFT with given SEV&CA parameters. The last column of table 5 shows that the frequency resolution reserve is enough to separate detection of total and difference frequencies.

For total and difference frequencies from table 5 a detailed EV analysis was carried out, its results given in table 4, similar to tables 2 and 3. Figures 8, 9 show some eigenvectors and their spectra whose data are given in table 4.
Having defined how many times the CI of selected eigenvectors exceeded the median value, we calculated the false alarm probability values which for certain TS were in the range from $1.9 \cdot 10^{-6}$ (Verkhneye Dubrovo station) to $4.9 \cdot 10^{-3}$ (Dusheti station).

The false detection probability for the fact of API-frequency modulation of components localized at BSS GW impact is negligible for all the analysed $E_z$ TS and does not exceed $1.5 \cdot 10^{-15}$. So, modulations should be considered an established fact.

Table 4. Coherence indexes and RMS values of eigenvectors spectrally localized at total and difference values of BSS GW impact frequencies and API frequency. Results of the observed $E_z$ time series analysis.

| Intermodulation frequency name | EV | EV CI | CI median | Median exceeding | Amplitude RMS value, V/m |
|---|---|---|---|---|---|
| VSU testing ground, $P_{fa}(17, 19)=9.8 \cdot 10^{-4}$ | | | | | |
| J1012+5307+ | 256; 257 | 71.9; 72;6 | 34.7 | 2 of 2 | 0.28; 0.28 |
| J1012+5307- | 173; 180 | 55.0; 44.5 | 34.7 | 2 of 2 | 0.42; 0.40 |
| J1537+1155+ | 620; 636 | 19.7; 24.3 | 30.4 | 0 of 2 | 0.13; 0.13 |
| J1537+1155- | 201; 202 | 211; 181 | 30.4 | 2 of 2 | 0.35; 0.35 |
| J1959+2048+ | 266; 267 | 194; 205 | 24.0 | 2 of 2 | 0.27; 0.27 |
| J1959+2048- | 576; 577 | 54.4; 55;6 | 24.0 | 2 of 2 | 0.15; 0.15 |
| J2130+1210+ | 421; 422 | 40.9; 49.9 | 27.3 | 2 of 2 | 0.19; 0.19 |
| J2130+1210- | 307; 308 | 44.4; 44.2 | 27.3 | 2 of 2 | 0.23; 0.23 |
| J1915+1606+ | 918; 919 | 143; 158 | 28.8 | 2 of 2 | 0.05; 0.05 |
| J1915+1606- | 300 | 82.6 | 28.8 | 1 of 1 | 0.24 |
| Voyeikovo, $P_{fa}(18, 138)=3.8 \cdot 10^{-6}$ | | | | | |
| J1012+5307+ | 340; 341 | 59.6; 146 | 34.7 | 2 of 2 | 0.38; 0.38 |
| J1012+5307- | 237 | 64.1 | 34.7 | 1 of 1 | 0.4 |
| J1537+1155+ | 438; 437 | 48.4; 39.8 | 30.4 | 2 of 2 | 0.33; 0.33 |
| J1537+1155- | 327; 328 | 71.8; 83.1 | 30.4 | 2 of 2 | 0.39; 0.39 |
| J1959+2048+ | 434 | 37.4 | 24.0 | 1 of 1 | 0.33 |
| J1959+2048- | 382; 430 | 64.6; 48.5 | 24.0 | 2 of 2 | 0.36; 0.36 |
| J2130+1210+ | 519; 520 | 118; 146 | 27.3 | 2 of 2 | 0.30; 0.30 |
| J2130+1210- | 474; 475 | 54/6; 88.1 | 27.3 | 2 of 2 | 0.31; 0.31 |
| J1915+1606+ | 472; 473 | 90.7; 94.3 | 28.8 | 2 of 2 | 0.31; 0.31 |
| J1915+1606- | 462; 463 | 64.9; 98 | 28.8 | 2 of 2 | 0.32; 0.32 |
| Dusheti, $P_{fa}(17, 20)=4.3 \cdot 10^{-3}$ | | | | | |
| J1012+5307+ | 324; 325 | 31.6; 56.4 | 34.7 | 2 of 2 | 0.46; 0.46 |
| J1012+5307- | 157; 158 | 121; 65.4 | 34.7 | 2 of 2 | 0.59; 0.59 |
| J1537+1155+ | 684; 685 | 79.2; 77.9 | 30.4 | 2 of 2 | 0.38; 0.38 |
| J1537+1155- | 408; 409 | 54.1; 49.7 | 30.4 | 2 of 2 | 0.43; 0.43 |
| J1959+2048+ | 279; 280 | 49.0; 59.6 | 24.0 | 2 of 2 | 0.49; 0.49 |
| J1959+2048- | 479; 480 | 39.6; 64.5 | 24.0 | 2 of 2 | 0.42; 0.42 |
| J2130+1210+ | 424; 425 | 23.4; 22.2 | 27.3 | 0 of 2 | 0.43; 0.43 |
| J2130+1210- | 466; 467 | 24.9; 32.2 | 27.3 | 1 of 2 | 0.42; 0.42 |
| J1915+1606+ | 528; 529 | 29.2; 35.5 | 28.8 | 2 of 2 | 0.41; 0.41 |
| J1915+1606- | 199; 201 | 49.8; 92.9 | 28.8 | 2 of 2 | 0.54; 0.54 |
| Verkhneye Dubrovo, $P_{fa}(19, 19)=1.910^{-6}$ | | | | | |
| J1012+5307+ | 343; 344 | 42.2; 51.1 | 34.7 | 2 of 2 | 0.34; 0.34 |
| J1012+5307- | 261; 262 | 138; 152 | 34.7 | 2 of 2 | 0.41; 0.41 |
| J1537+1155+ | 433 | 42.2 | 30.4 | 1 of 1 | 0.3 |
| J1537+1155- | 418; 419 | 65.3; 112 | 30.4 | 2 of 2 | 0.3; 0.3 |
| J1959+2048+ | 500; 501 | 58.9; 66.9 | 24.0 | 2 of 2 | 0.27; 0.27 |
| J1959+2048- | 378; 383 | 42.7; 47.6 | 24.0 | 2 of 2 | 0.33; 0.32 |
| J2130+1210+ | 459; 460 | 84.1; 98.7 | 27.3 | 2 of 2 | 0.29; 0.29 |
| J2130+1210- | 455; 468 | 61.9; 42.3 | 27.3 | 2 of 2 | 0.29; 0.28 |
| J1915+1606+ | 530; 531 | 102; 47.5 | 28.8 | 2 of 2 | 0.27; 0.26 |
| J1915+1606- | 490; 491 | 127; 116 | 28.8 | 2 of 2 | 0.27; 0.27 |
| For all the four time series, $P_{fa}(71, 76)=1.2 \cdot 10^{-15}$ | | | | | |

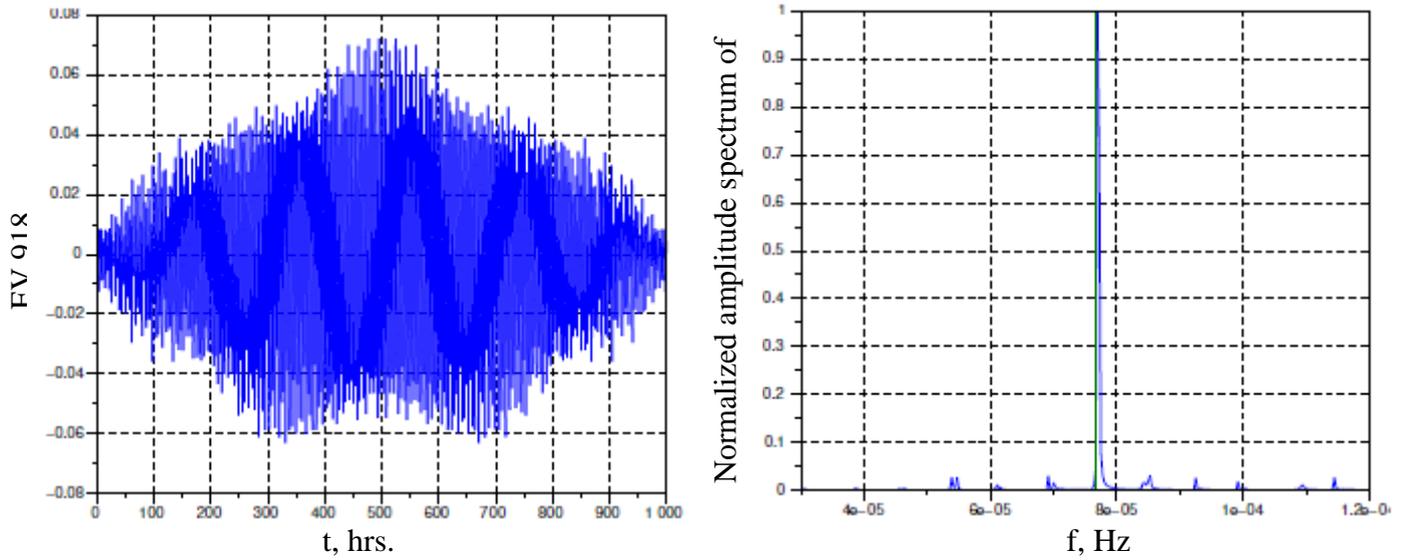
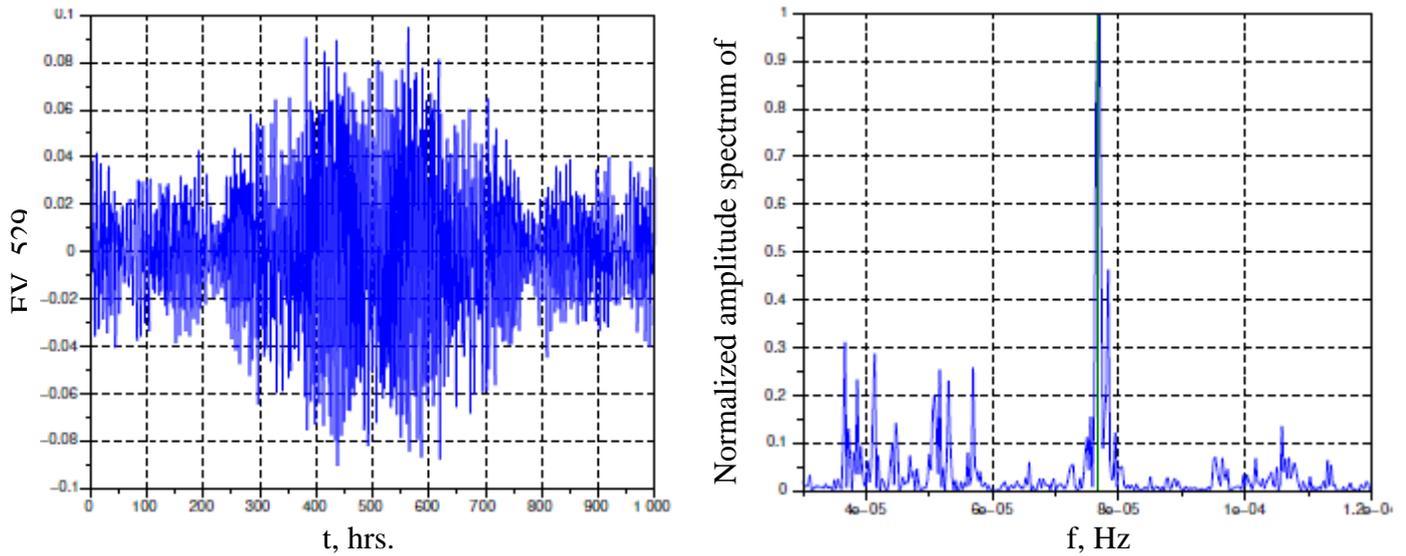
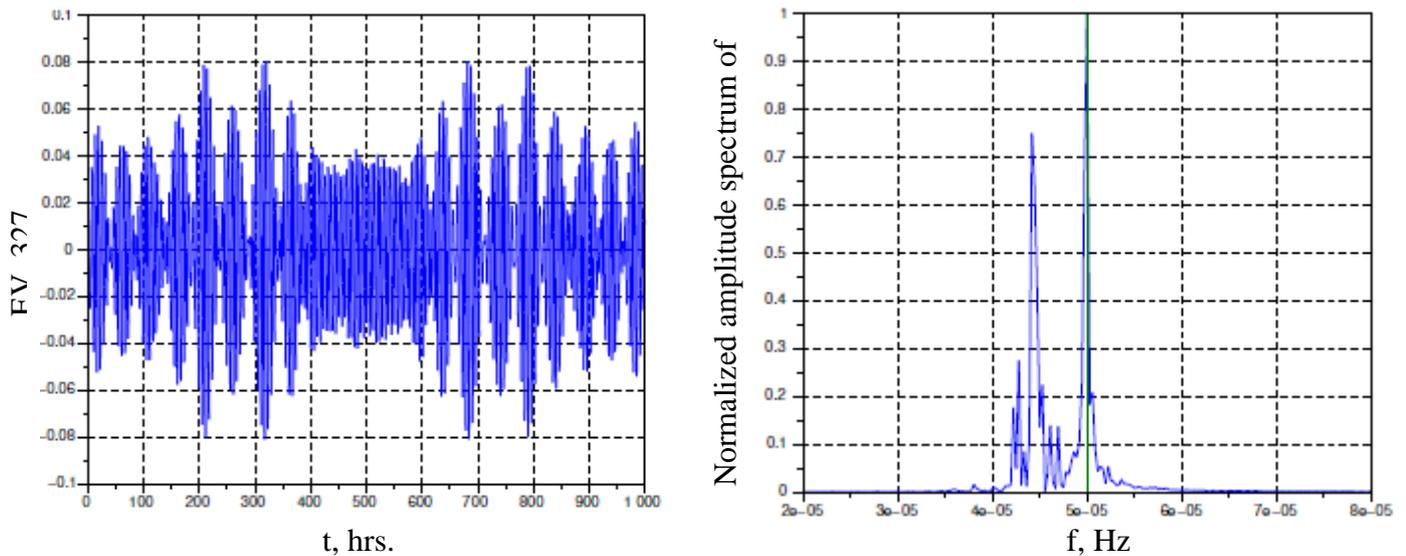

Fig.8. Eigenvectors localized at intermodulation frequencies of BSS GW impact and API (left) and the EV normalized amplitude spectra (right). Continuous vertical line at the spectrum plots relates to the considered frequency.

Total frequency of J1012+5307 GW impact and API, $E_z$ TS, data from VSU testing ground

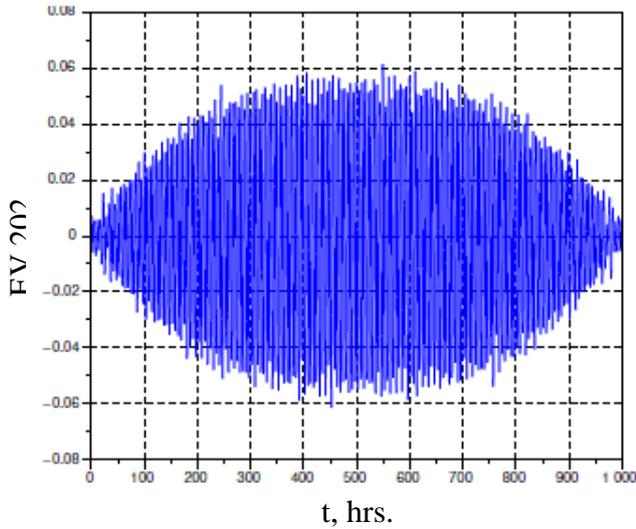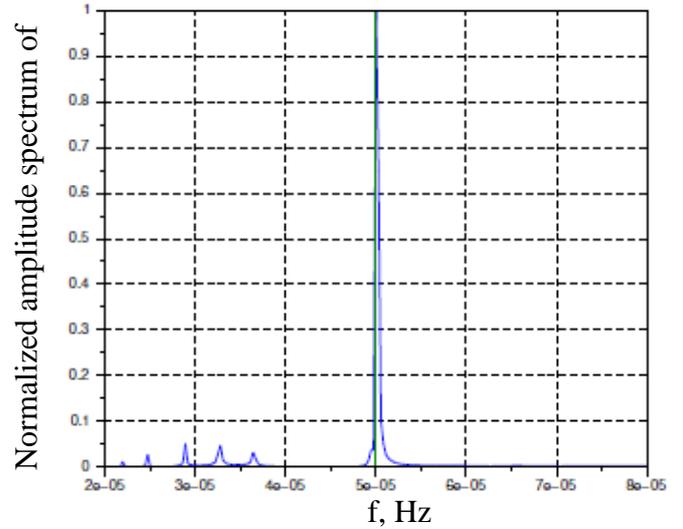

Total frequency of J1012+5307 GW impact and API, $E_z$ TS, data from Dusheti station

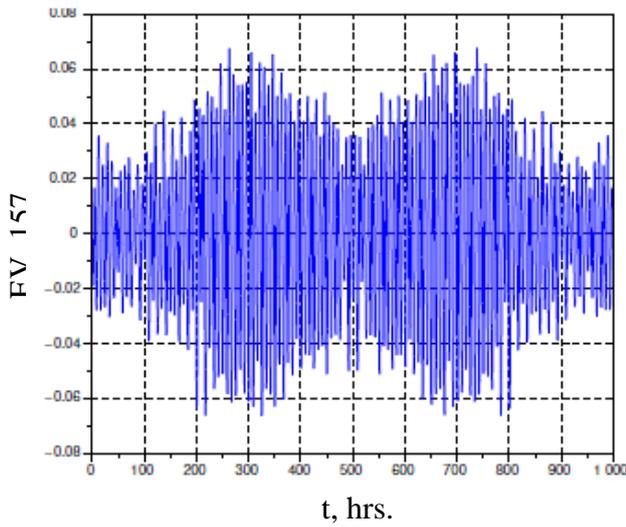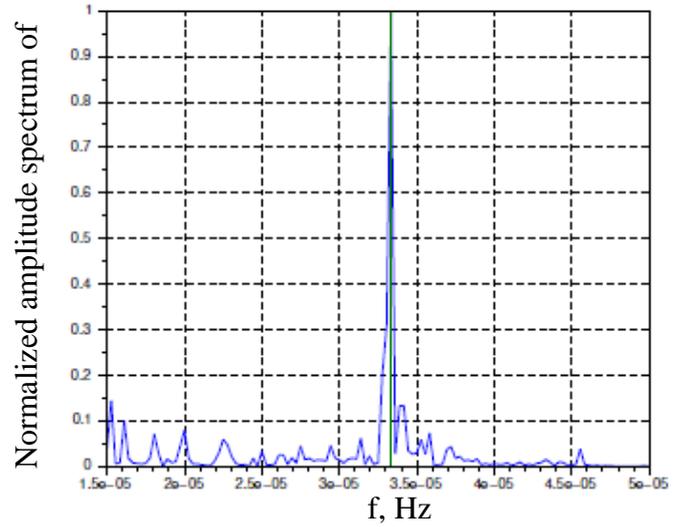

Total frequency of J1012+5307 GW impact and API, $E_z$ TS, data from Verkhneye Dubrovo station

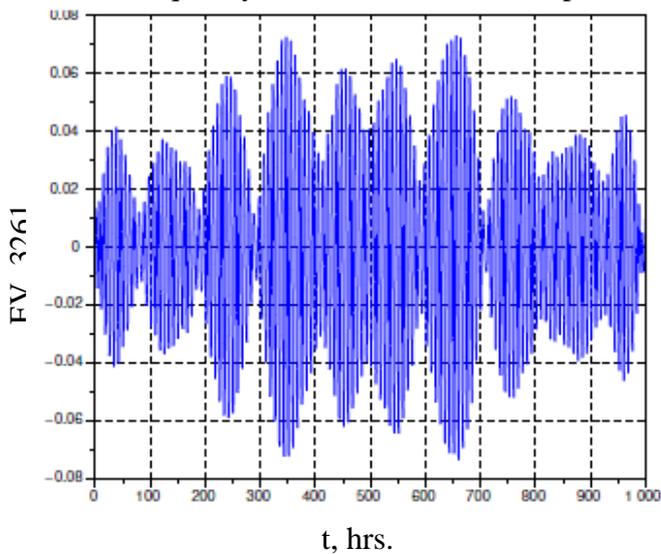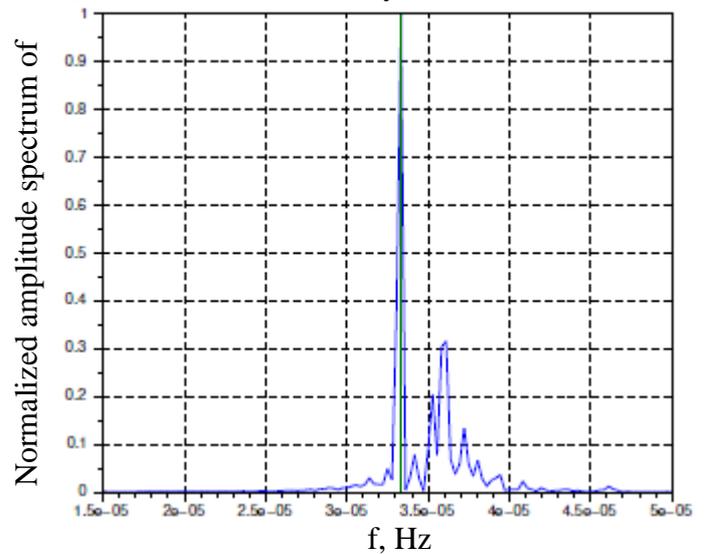

Fig.9. Eigenvectors localized at intermodulation frequencies of BSS GW impact and API (left) and the EV normalized amplitude spectra (right). Continuous vertical line at the spectrum plots relates to the considered frequency.

Total frequency of tide $S_3$ and API, $E_z$ TS, data from Voyeikovo station

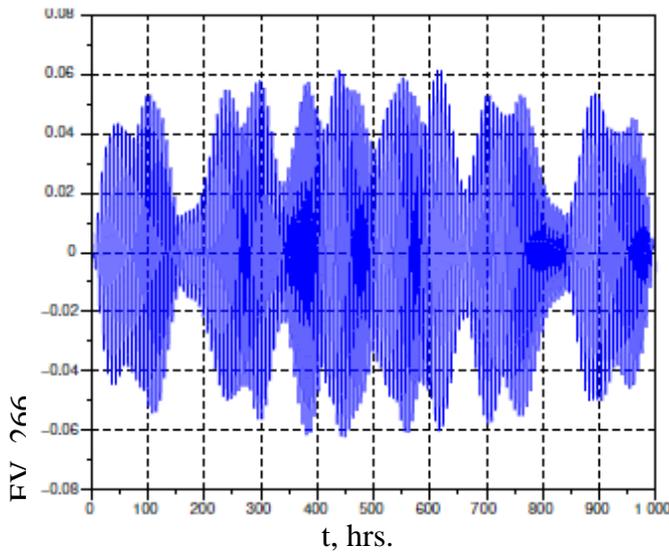 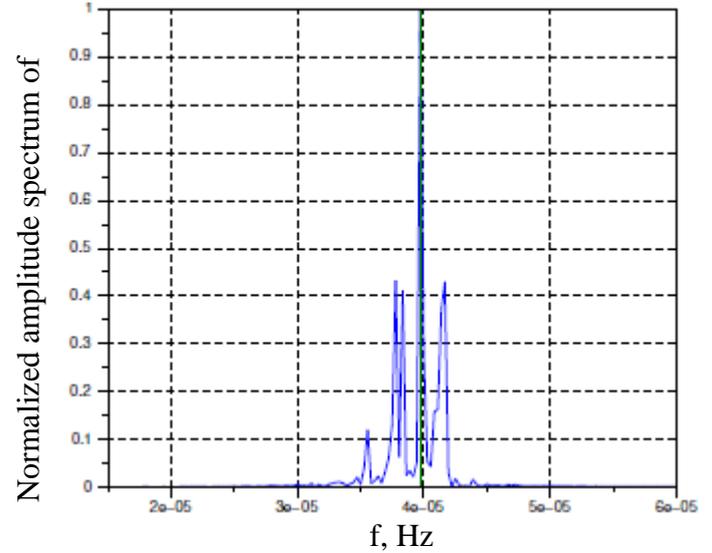

Difference frequency of tide $S_3$ and API, $E_z$ TS, data from Voyeikovo station

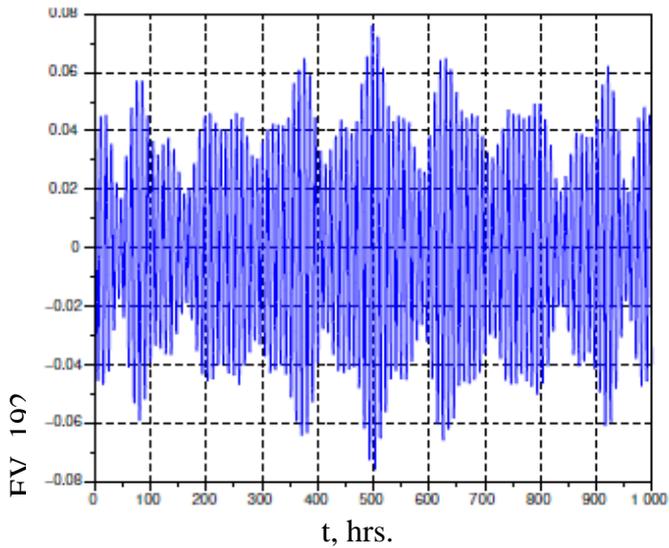 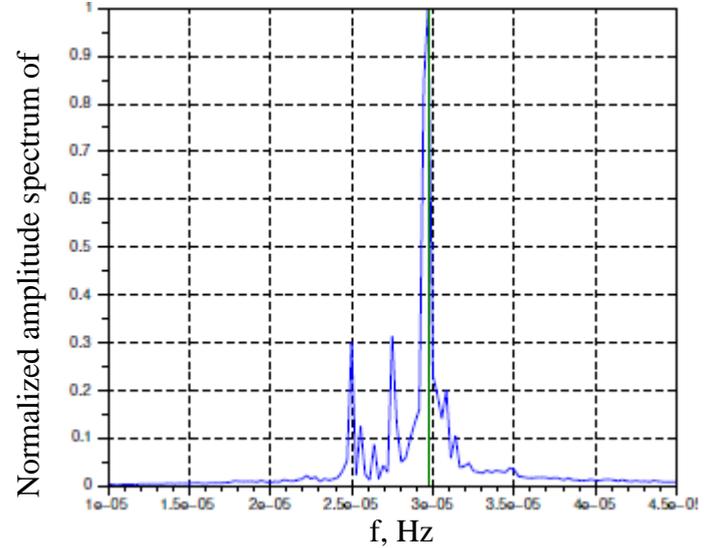

Difference frequency of lunar tide $J_1$ and API, $E_z$ TS, data from Voyeikovo station

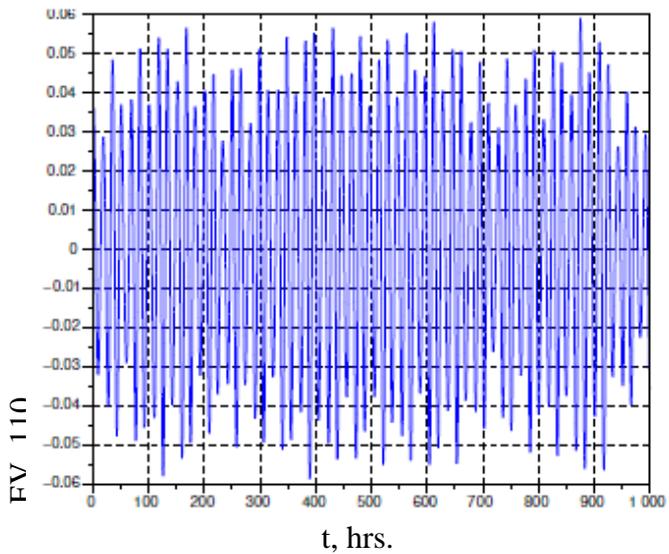 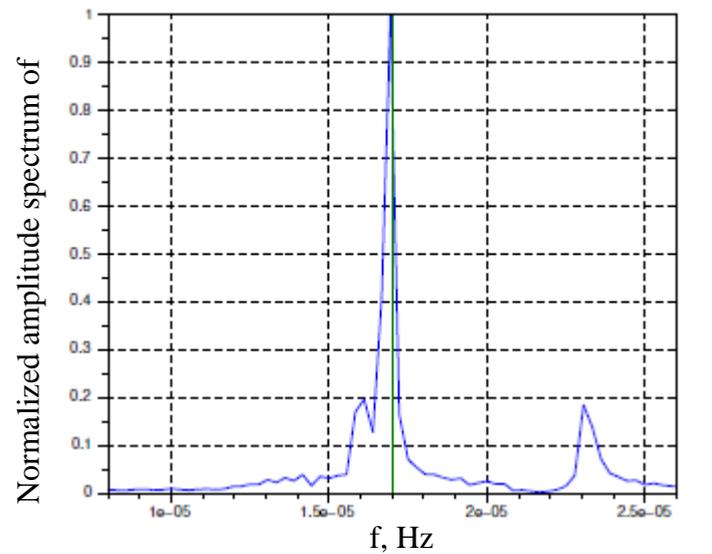

Fig10. Eigenvectors localized at intermodulation frequencies of lunar/solar tides and API (left) and the EV normalized amplitude spectra (right). Continuous vertical line at the spectrum plots relates to the considered frequency.

Table 5. Considered total (+) and difference (-) frequencies of BSS GW impact and API.

| Intermodulation source considered | | Source nearest by frequency | | $\|f_{cons} - f_{nearest}\|$ |
| --- | --- | --- | --- | --- |
| Name | $f_{cons}$, $10^{-5}$ Hz | Name | $f_{nearest}$, $10^{-5}$ Hz | $\Delta f$ |
| J1012+5307+ | | $S_4$ | | 11 |
| J1012+5307- | | $S_3$ | | 6 |
| J1537+1155+ | | J1959+2048 | | 3 |
| J1537+11 55- | | $S_4$ | | 14 |
| J1959+2048+ | | J2130+1210- | | 6 |
| J1959+2048- | | J1537+1155 | | 3 |
| J2130+1210+ | | J1915+1606 | | 9 |
| J2130+1210- | | J1915+1606- | | 10 |
| J1915+1606+ | | J2130+1210+ | | 10 |
| J1915+1606- | | J2130+1210 | | 9 |

The effect of API-frequency modulation detected for components having BSS GW impact frequencies makes us to raise a question of how wide –spread are API-frequency amplitude modulations among quasi-periodic processes observed in the electric field vertical component of the atmosphere boundary layer. Are spread the hypothesis of these modulations being wide-spread gets preliminary confirmation at figure 10 demonstrating some eigenvectors localized at total and difference frequencies for one lunar tide and one solar tide. The figure shows spectral localization at total and difference frequencies, so the hypothesis has a reason to be considered. Confirming the hypothesis, though, is a subject of a separate work.

The results of this work were obtained using the authors' registered methods and software [13, 14].

**Conclusions**

1. Time series of $E_z$ vertical component of the Earth atmosphere electric field boundary layer have components localized at the frequencies of binary stars gravity-wave impact or at the frequency of axion-photon interaction. It was demonstrated that these components are non-coherent. This is the reason why increasing the analysis time range leads to decreasing of spectral estimates got by means of conventional quadrature scheme. So, the components localized at the named frequencies cannot be detected using quadrature scheme.

2. Using signal eigenvectors and components analyser [3] (SEV&CA) made it possible to detect with high reliability the non-coherent components localized at the frequencies of binary stars gravity-wave impact. For all the analysed $E_z$ time series the false detection probability is not more than $10^{-9}$.

3. Using SEV&CA made it possible to detect $E_z$ non-coherent components spectrally localized at axion-photon frequency of $5 \cdot 10^{-6}$ Hz. For all the analysed $E_z$ time series the false detection probability is not more than 0.06.

4. We've detected that the amplitude of non-coherent components spectrally localized at binary stars gravity-wave impact is modulated by the frequency equal to that of axion-photon interaction. For all the analysed $E_z$ time series the false detection probability of these modulations is negligible and does not exceed $1.5 \cdot 10^{-15}$. The detection of these modulations confirms with high reliability the fact of axion-photon interaction.

5. The hypothesis was proposed and preliminary confirmed that components modulated by axion-photon interaction frequency (5·10-6 Hz) are wide-spread among the quasi-periodic components of the electric field vertical component in the Earth atmosphere boundary layer.

**Abbreviations**

| API | axion-photon interaction |
|---|---|
| BSS | binary star system, binary star |
| CI | coherence index |
| EV | eigenvector |
| FFT | Fast Fourier transform |
| GW | gravity-wave |
| GWN | Gaussian white noise |
| ILF | infra-low frequency |
| NEVS | normalized eigenvalues spectrum |
| SEV&CA | signal eigenvectors and components analyser, eigenoscope |
| TS | time series |